\documentclass{elsart}
\usepackage{natbib}
\usepackage{amsmath}
\usepackage{amssymb}
\usepackage{type1cm}
\usepackage{graphicx}
\usepackage{amssymb}

\begin{document}

\begin{frontmatter}


\title{On the Explosion Mechanism of SNe Type Ia}
\author{M.\ Reinecke},
\ead{martin@mpa-garching.mpg.de}
\author{J.\ C.\ Niemeyer}
\and
\author{W.\ Hillebrandt}
\address{Max-Planck-Institut f\"ur Astrophysik,
 Karl-Schwarzschild-Str.\ 1, 85741 Garching, Germany}

\begin{abstract}
In this article we discuss the first simulations of two- and three-dimensional
Type Ia supernovae with an improved hydrodynamics code. After describing
the various enhancements, the obtained results are compared to those of
earlier code versions, observational data and the findings of other researchers
in this field.
\end{abstract}


\end{frontmatter}

\section{Introduction}
\label{}

The most generally accepted scenario for the explosion of a Type Ia supernova
consists of a thermonuclear reaction front propagating through a white dwarf
of Chandrasekhar mass, which is subsequently disrupted. Very early on it was
discovered that this combustion process cannot start out as a supersonic
detonation wave, but has to burn at subsonic speeds at least during the early
stages, in order to explain the abundances observed in SN Ia remnants
(see, e.g., \citealt{hillebrandt-niemeyer-2000}).

This fact complicates the numerical modeling of such a process mainly for two
reasons:
\begin{itemize}
  \item The buoyancy of the hot ashes near the stellar center leads to
  the development of hydrodynamic instabilities on many scales, which
  have a significant influence on the explosion dynamics, mainly by increasing
  the total flame surface and therefore the energy generation rate. Only a small
  part of the relevant scales can be resolved, so that a large region of
  scale space must be described by numerical turbulence models.
  \item Since the resulting turbulent flow pattern inside the exploding white
  dwarf is inherently three-dimensional, reliable quantitative results can only
  be obtained by resource-intensive 3D numerical calculations. Two-dimensional
  simulations, which imply axial symmetry conditions, may exhibit a different
  turbulent energy and velocity cascade and also underestimate the total
  burning surface. Consequently, they can only provide qualitative results and
  serve for validation of the numerical models.
\end{itemize}

In the following we shortly describe the numerical techniques employed in our
SN Ia code and discuss the results of 2D and 3D calculations and their
relation to observational data.

\section{Numerical methods}
The ensemble of numerical techniques used for the presented simulations is
based on those described by \cite{reinecke-etal-99a}; however, several
physical and numerical corrections and improvements were implemented in the
meantime:
\begin{itemize}
  \item The original minimalistic combustion scheme (one-step reaction from
    carbon/oxygen to nickel) was extended considerably. In the new version
    the composition of the ashes depends on the density of the fuel,
    i.e.\ a NSE-like mixture of nickel and $\alpha$-particles is synthesized
    at high densities and $^{24}$Mg (representing intermediate-mass elements)
    is produced at lower densities. Additionally, the composition in
    the Ni/$\alpha$ material tracks the NSE equilibrium during the expansion
    of the star, leading to delayed energy release.
  \item An incorrect numerical factor was found and corrected in the calculation
    of the source terms of the two-dimensional sub-grid model for the flame
    propagation speed. This mistake did not have a significant effect on the
    results published earlier (see, e.g., \citealt{reinecke-etal-99b}),
    but becomes noticeable if resolution studies
    are performed (see the next section).
  \item All numerical models were extended to allow three-dimensional
    simulations in Cartesian coordinates. Since 3D calculations are much more
    resource-intensive than 2D runs with comparable resolution, a
    parallelization of the complete numerical code was also necessary in order
    to run it on massively parallel computers.
\end{itemize}

\section{SN calculations}
Different series of simulations were performed to check the numerical
reliability of the employed models and to compare two- and three-dimensional
explosions.

\subsection{Resolution study}
A crucial test for the validity of the models for the unresolved scales
(in this case the flame and subgrid models) is to check the dependence of
integral quantities, like the total energy release of the explosion, on the
numerical grid resolution. Ideally, there should be no such dependence,
indicating that all effects on unresolved scales are accurately modeled.

\begin{figure}[tbp]
  \centerline{\includegraphics[width=0.7\textwidth]{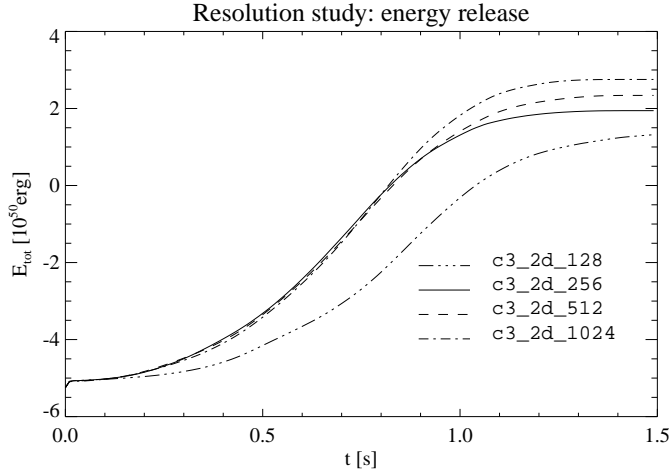}}
  \caption{Time evolution of the total energy for identical initial conditions
    at different resolutions. While model c3\_2d\_128 is clearly under-resolved,
    the other simulations agree very well, at least in the early and
    intermediate stages.}
  \label{e_compare}
\end{figure}

Figure \ref{e_compare} shows the energy evolution of a centrally ignited white
dwarf. The only difference between the simulations is the central grid
resolution,
which ranges from $2\cdot10^6$cm (model c3\_2d\_128) down to $2.5\cdot10^5$cm
(model c3\_2d\_1024). Model c3\_2d\_128 is obviously under-resolved, but the
results of the other calculations are in good agreement, with exception of
the last stages, where the flame enters strongly non-uniform regions of the
grid.

So far, this kind of parameter study could only be performed in two dimensions,
because of the prohibitive cost of very highly resolved 3D simulations.
Nevertheless the results suggest that a resolution of $\approx10^6$cm should
yield acceptable accuracy also in three dimensions. 

\subsection{Comparison of 2D and 3D simulations}
\begin{figure}[tbp]
  \centerline{\includegraphics[width=\textwidth]{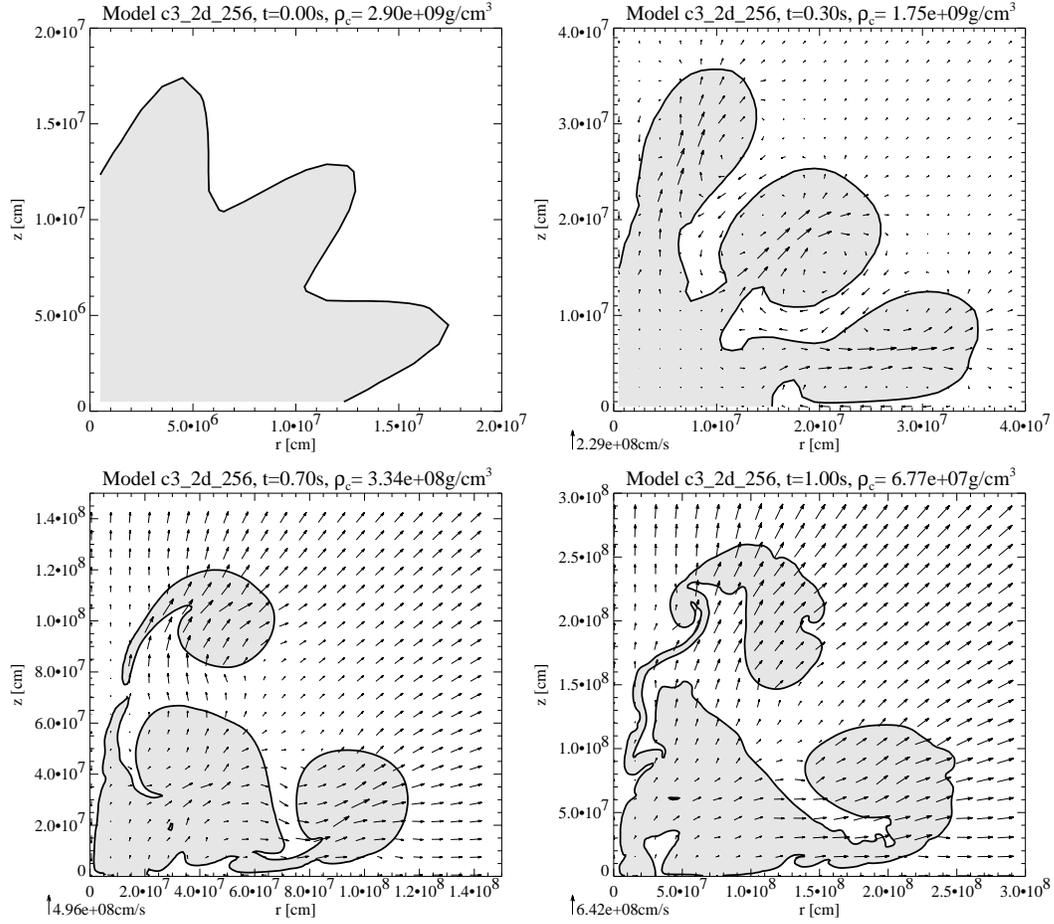}}
  \caption{Burning front geometry evolution of model c3\_2d\_256.}
  \label{front2d}
\end{figure}

\begin{figure}[tbp]
  \centerline{\includegraphics[width=\textwidth]{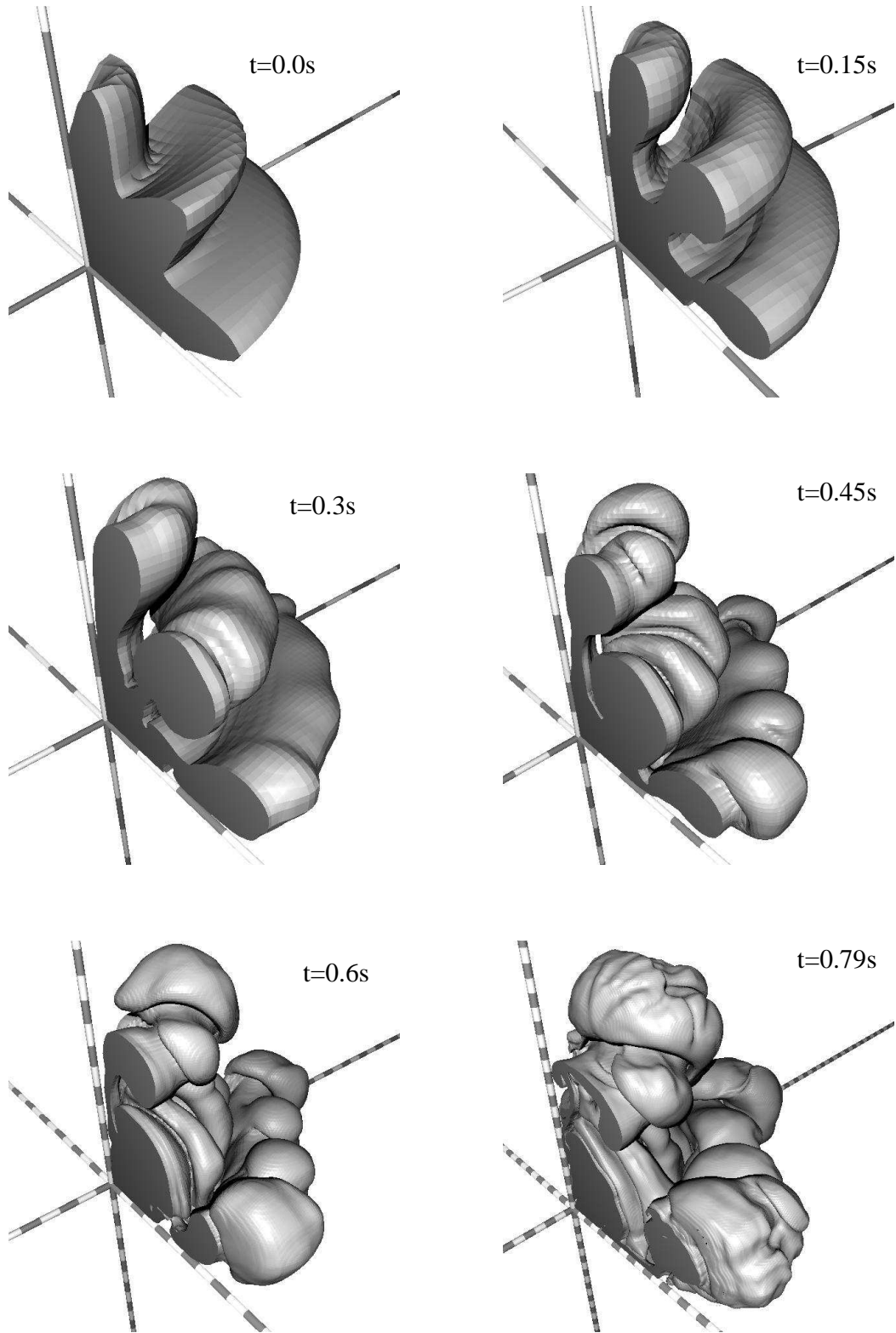}}
  \caption{Burning front geometry evolution of model c3\_3d\_256.
           One ring on the axes corresponds to $10^7$cm.}
  \label{front3d}
\end{figure}

\begin{figure}[tbp]
  \centerline{\includegraphics[width=0.7\textwidth]{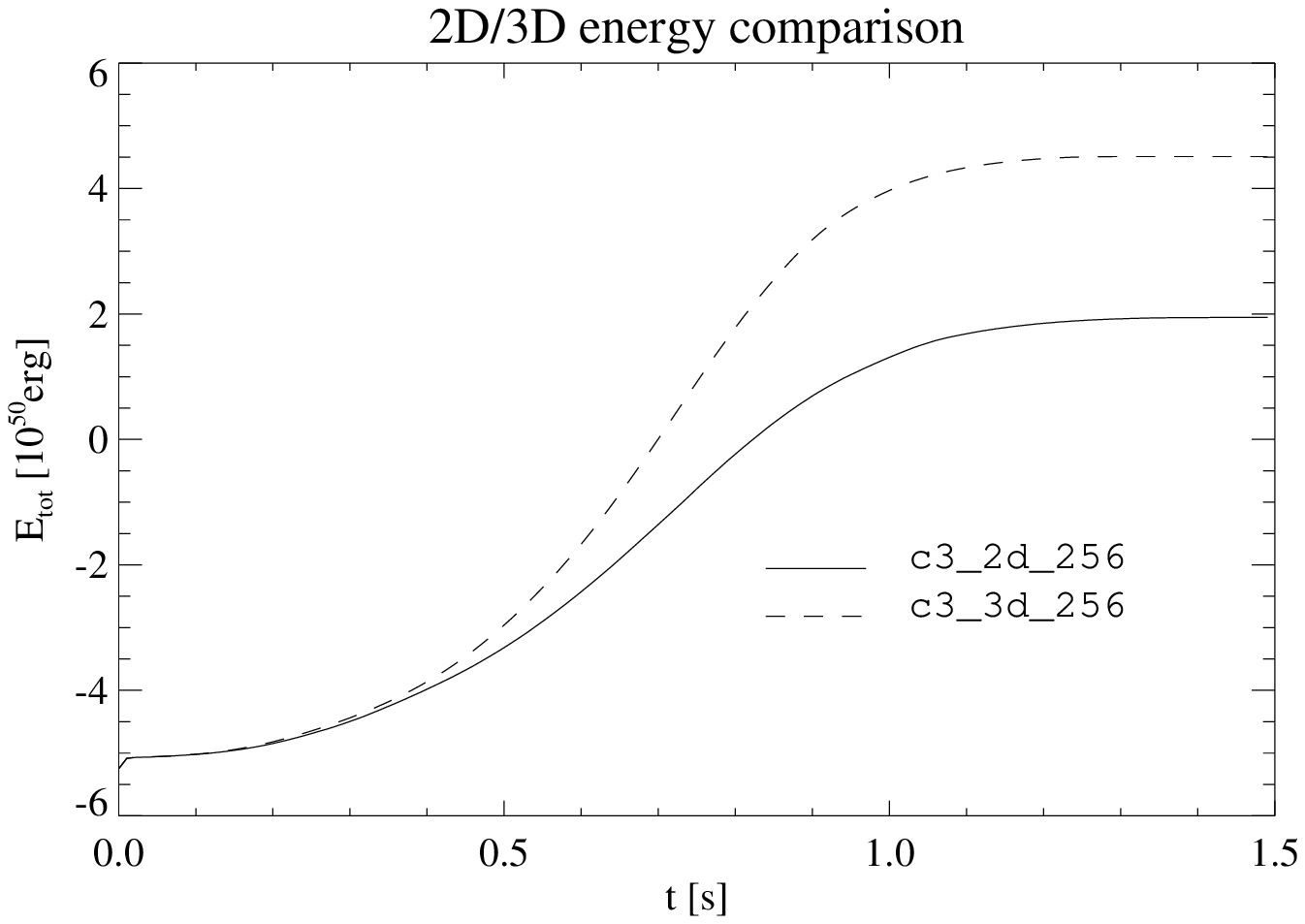}}
  \caption{Comparison of the energy evolution for identical initial
           conditions in two and three dimensions.}
  \label{comp2d3d}
\end{figure}

In order to investigate the fundamental differences between two- and
three-dimensional simulations, a 2D and a 3D model with identical initial
conditions and resolution was calculated.
Figures \ref{front2d} and \ref{front3d} show
snapshots of the flame geometry at various explosion stages; the energy
evolution of both models is compared in figure \ref{comp2d3d}.


It is evident that both simulations evolve nearly identically during the first
few tenths of a second, as was expected. This is a strong hint that no errors
were introduced into the code during the enhancement of the numerical models
to three dimensions. At later times, however, the 3D calculation develops
instabilities in the azimuthal direction, which could not form in 2D because
of the assumed axial symmetry. As a consequence the total burning surface
and the energy generation rate is increased, resulting in a higher overall
energy release.

\section{Conclusions}
Table \ref{burntable} lists the energy releases, as well as the masses of
intermediate elements and nickel for the three-dimensional simulations
performed up to date. These results (in contrast to the 2D models, which
give too weak explosions) agree fairly well with the energies
and nickel masses derived from observations \citep{contardo-etal-00}.
The results exhibit noticeable scatter for different initial conditions,
i.e.\ the location of the flame at the beginning of the thermonuclear runaway
seems to have an important influence on the supernova energetics. Since very
little is known about this parameter, one has to investigate, as a next step,
the secular pre-ignition evolution of the white dwarf, which finally determines
the ignition conditions.

\begin{table}[htbp]
  \centerline{
    \begin{tabular}{|l|c|c|c|}
    \hline
    model name & $m_{\text{Mg}}$ [$M_\odot$]&$m_{\text{Ni}}$ [$M_\odot$]&$E_{\text{nuc}}$ [$10^{50}$\,erg\vphantom{\raisebox{2pt}{\large A}}] \\
    \hline
    c3\_3d\_256 & 0.177 & 0.526 & 9.76 \\ \hline
    b5\_3d\_256 & 0.180 & 0.506 & 9.47 \\ \hline
    b9\_3d\_512 & 0.190 & 0.616 & 11.26\phantom{0} \\ \hline
    \end{tabular}
  }
  \caption{Overview over element production and energy release of the
   supernova simulations performed with the presented code. In contrast
   to the centrally ignited model c3\_3d\_256, models b5\_3d\_256 and
   b9\_3d\_512 were ignited in 5 resp.\ 9 spherical bubbles, which were
   randomly distributed in the vicinity of the center. In addition, model
   b9\_3d\_512 was calculated at higher resolution.}
  \label{burntable}
\end{table}

In addition to the good agreement with observations, it should be noted
that our results also are quite similar to those obtained by
\cite{khokhlov-00}, who approached the task of modeling the deflagration
in a SN Ia in a quite different way. This could indicate that our
understanding of all relevant physical processes is now sufficient to
devise correct numerical models and converge towards the real solutions.


\bibliographystyle{elsart-harv}
\bibliography{refs}

\end{document}